\documentclass[aps,twocolumn,prd,showpacs,nofootinbib]{revtex4}
\usepackage{amsmath}
\usepackage{graphicx}
\usepackage{dcolumn}
\usepackage{bm}
\usepackage{amssymb}
\usepackage{latexsym}

\def\be{\begin{equation}}
\def\ee{\end{equation}}
\def\ba{\begin{eqnarray}}
\def\ea{\end{eqnarray}}

\begin{document}

\title{Design of a Cyclic Multiverse }

\author{Yun-Song Piao}

\affiliation{College of Physical Sciences, Graduate School of
Chinese Academy of Sciences, Beijing 100049, China}

\begin{abstract}

Recently, it has been noticed that the amplification of the
amplitude of curvature perturbation cycle by cycle can lead to a
cyclic multiverse scenario, in which the number of universes
increases cycle by cycle. However, this amplification will also
inevitably induce either the ultimate end of corresponding cycle,
or the resulting spectrum of perturbations inside corresponding
universe is not scale invariant, which baffles the existence of
observable universes. In this paper, we propose a design of a
cyclic multiverse, in which the observable universe can emerges
naturally. The significance of a long period of dark energy before
the turnaround of each cycle for this implementing is shown.

\end{abstract}

\maketitle


Recently, a cosmological cyclic scenario, in which the universe
experiences the periodic sequence of contractions and expansions
\cite{Tolman}, has been rewaked \cite{STS}, and brought the
distinct insights into the origin of observable universe. There
has been lots of studies for cyclic or oscillating universe
\cite{BD},\cite{KSS},\cite{Piao04},\cite{Lidsey04},\cite{CB},\cite{Xiong},\cite{Xin},\cite{LS},\cite{Biswas},\cite{Biswas1},
also \cite{NB} for a review. However, the global configuration of
cyclic universe is actually more complex than imagined. In cyclic
universe the amplitude of the curvature perturbation on super
horizon scale is generally increasing during the contraction of
each cycle, while is nearly constant during the expansion of each
cycle. Thus the net result is that the amplitude of perturbation
is amplified, which occurs cycle by cycle,

Recently, it has been argued that the amplification of the
amplitude of curvature perturbation cycle by cycle can lead to a
cyclic multiverse scenario \cite{Piao0901}\footnote{Here, the
cyclic multiverse means that there are many independent universes
in each cycle or each spacelike slice. While in usual cyclic
universe there is only single universe in each cycle or each
spacelike slice, which can be regarded as a multiverse only when
we count it along the time sequence.}, in which the universe
proliferates, i.e. the number of universes increases cycle by
cycle. There this was illustrated by including a contracting phase
with $w\simeq 0$ in each cycle of a cycle universe. In principle,
since shortly after the end of each cycle the curvature
perturbation on super horizon scale can be amplified to order one,
the universe will be inevitably separated into lots of parts
independent of one another after each cycle,
each of which actually corresponds to a new universe and
independently evolves up to succedent cycle, and then proliferates
again \footnote{In \cite{Feng04}, an eternal expanding recurrent
universe, in which $h$ oscillates periodically while $a$ expands
all along, has been proposed phenomenologically, also latest
\cite{IBF}, which can be implemented by appealing to a phantom
component with $w<-1$. In this scenario, we live in one period of
cycling, while the present acceleration with $w<-1$ is just a
start of the phantom inflation \cite{PZ},\cite{PZ1} for the next
period of cycling. Similarly, a multiverse scenario might be
obtained here, which will be explored.}. This result in some sense
incorporates the second law of thermodynamics in such a fashion
that the increase of total entropy in consecutive cycles is
explained as or replaced with the increase of the number of new
universes, which has been discussed in \cite{Piao0901} in
detailed.

The cyclic multiverse scenario might be significant for
understanding the origin of observable universe. However,
generally the amplitude of perturbation modes that enter into the
horizon during the expansion in previous cycle and then leave it
during the contraction in current cycle is generally larger than
that induced by the quantum fluctuation of background field in
current cycle. This will inevitably lead to either the ultimate
end of corresponding cycle, or the resulting spectrum of
perturbations inside each universe is not scale invariant even if
they leave the horizon during the contraction with $w\simeq 0$,
which baffles possible existence of observable universes in this
scenario. We, in this paper, will discuss this problem in detail
and then design a possible solution.
We begin with an illustration of cyclic universe model and the
review of the evolution of perturbation in a cyclic universe with
the contraction with $w\simeq 0$. There have been lots of studies
of bounce cosmology \cite{NB}. Recently, the nonsingular bounce
has been implemented in nonlocal higher derivative theories of
gravity \cite{BKM}, which is ghostfree, while here that provided
by us is only a simple example serving the purpose of
illustration. We introduce a normal field $\varphi$ with its
potential ${{M}^2\varphi^2}-\Lambda_*$, and a field $\chi$ with
negative ${\dot\chi}^2$, which plays a crucial role in giving a
nonsingular bounce \cite{Cai0704}. $\Lambda_*$ is a small positive
constant. Thus the minimum of potential is negative, which is
responsible for the turnaround of cyclic universe. We show such an
illustration of cyclic universe in Fig.1, in which ${M}=0.9$,
$\Lambda_*=10^{-10}$, the initial value $10^4\varphi_0=5{M}$ and
$10^4\chi_0=3{M}$ are used and $M_P=1$ is set. It seems that this
model suffers from the quantum instabilities and inconsistencies
\cite{BKM}. However, it can be expected that this effective
description is only an approximation of a fundamental theory with
the UV completion below certain physical cutoff, while the
appearance of ghost terms or the quantum instabilities is only an
artefact of this approximation. In string theory, a slowly
decaying D3 brane can be depicted by an open string tachyon mode,
which is described within a open string field theory. This will
bring a nonlocal polynomial interaction. In this case, the
behavior of open string tachyon can be effectively simulated by a
ghost field, e.g.\cite{AJK},\cite{AKV}, as used here. However, the
string theory is expected to be UV complete.

\begin{figure}[t]
\begin{center}
\includegraphics[width=7.5cm]{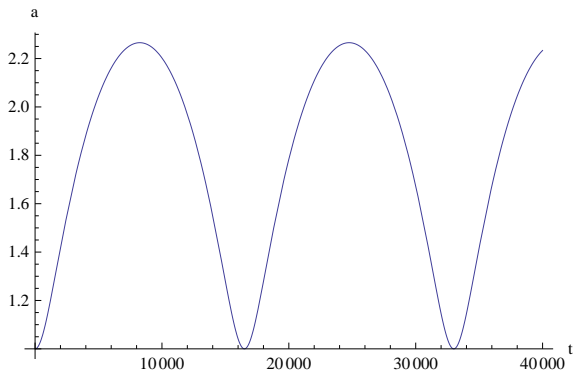}
\caption{ A simple model of cyclic universe}
\end{center}
\end{figure}


We will regard the beginning of the contracting phase as the
beginning of a cycle, in each cycle the universe will orderly
experience the contraction, bounce, and expansion, and then arrive
at the turnaround, which signals the end of a cycle. The motive
equation for perturbation in the momentum space is \be
u_k^{\prime\prime} +\left(k^2-{z^{\prime\prime}\over z}\right) u_k
= 0 ,\label{uk}\ee where $u_k$ is related to the curvature
perturbation $\zeta$ by $u_k \equiv z\zeta_k$
\cite{Muk},\cite{KS}, and the prime denotes the derivative with
respect to the conformal time $\eta$, and $z={{\varphi^\prime}/
h}$, where $h$ is the Hubble parameter and $\varphi$ is the
background field.
When $w\simeq 0$, $z\sim \eta^2$. Thus we have
${z^{\prime\prime}\over z} \sim {2\over \eta^2}$, which is
actually the same as that in inflation, in which $a\sim {1\over
\eta}$ leads $z\sim {1\over \eta}$ and thus
${z^{\prime\prime}\over z} \sim {2\over \eta^2}$. This gives the
spectrum $n_s\simeq 1$ is scale invariant, as has been shown in
\cite{Wands99, FB}, also see
\cite{PP},\cite{Wands09},\cite{Cai0810} and earlier \cite{S} for
tensor perturbation. Thus the amplitude of curvature perturbation
is given by \be {\cal P}_{\zeta}^{1/2}\simeq
k^{3/2}\left|{u_k\over z}\right|\simeq {h_e\over m_p},
\label{p1}\ee where we neglected the factor with order one, and
$h_e$ is determined by the energy scale $\rho_e$ at the end time
of contracting phase, $h_e\simeq {\sqrt{\rho_e}\over m_p}$. This
amplitude of perturbation spectrum is actually determined by the
increasing mode of metric perturbation $\Phi$ during the
contraction, which enters into the constant mode of $\zeta$ or
$\Phi$ after the bounce by $k^2$ order of $\zeta$.
The amplitude of curvature perturbation during the contraction is
increased, up to the end of contracting phase in corresponding
cycle, while during the expansion it becomes constant on super
horizon scale. Thus for a cycle the net result is the amplitude of
curvature perturbation on super horizon scale is amplified, which
is inevitable here.

\begin{figure}[t]
\begin{center}
\includegraphics[width=7.5cm]{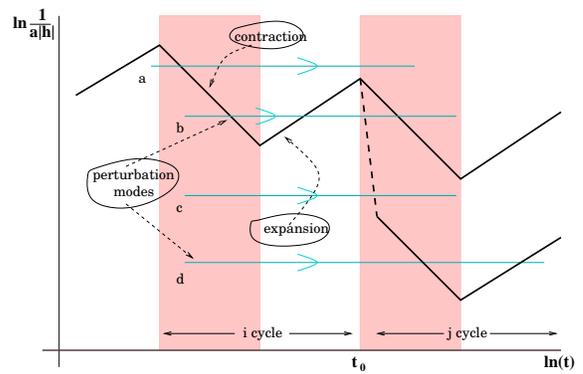}
\caption{In a cyclic universe, the sketch of $\ln{({1\over
a|h|})}$ vs. $\ln(t)$. The adjacent cycles have been signed as $i$
and $j$ cycles, respectively. The blue lines denote the evolutions
of perturbation modes, and the red regions denote the contracting
phases in each cycle. The dashed line denotes the period of dark
energy domination, and the following solid line denotes the
evolution after it. $t_0$ is the turnaround epoch. In principle,
at the bounce and turnaround points, $h = 0$, thus there is a
divergence for ${1\over a|h|}$ , which is not plotted here and
also Fig.3 and 4. However, the discussions are not affected by
this neglect. }
\end{center}
\end{figure}





We then consider the evolution of perturbation, which is generated
in previous cycle, entering into current cycle. We regard adjacent
cycles as $i$ and $j$ cycles for convenience, respectively. The
solution of metric perturbation $\Phi$ is $\Phi\simeq
\Phi_C+\Phi_S{h/a}$. When $w\simeq 0$, during the contraction the
constant mode $\Phi_C\sim \sqrt{k^{3}}$ and the increased mode
$\Phi_S\sim {1/\sqrt{k^{7}}}$. The solutions of $\zeta$ is
generally $ \zeta\simeq {\Phi_C}+ k^2{\Phi_S}f(\eta)$ \cite{FB},
where \be f(\eta)=\int {d\eta\over z^2}\sim {1\over \eta^3}\sim
{h} \label{f}\ee for $w\simeq 0$, since $z\sim \eta^2$ and $h\sim
1/\eta^3$. There is not bounce, but the contraction phase after
$t_0$. Thus $\zeta$ is dominated by its increasing mode $\Phi_S$,
i.e. $\zeta\simeq k^2{\Phi_S}f(\eta)\sim k^{-3/2} h$. Thus ${\cal
P}^{1/2}_{\zeta}\simeq k^{3/2}|\zeta|\sim h$ is scale invariant.
In $j$ cycle, since at the beginning time $t_0$, $h=h_0$, see
Fig.2, and the initial value of perturbation is ${\cal
P}_{\zeta(i)}^{1/2}$, the amplitude of perturbation mode all along
on super horizon scale, which is generated during the contraction
in $i$ cycle and can not reenter into the horizon during the
expansion of $i$ cycle, see `a' mode in Fig.1, is given by \be
{\cal P}_{\zeta(ji)}^{1/2}= {\cal
P}_{\zeta(i)}^{1/2}\left({h_{j}\over h_0}\right)= {\cal
P}_{\zeta(i)}^{1/2}e^{3{\cal N}_{j}}, \label{p3}\ee where the
subscript $i$ and $j$ denote the quantities in corresponding
cycles, respectively, and ${\cal N}_{j}={1\over 3}\ln{({h_{j}\over
h_0})}$ is the efolding number that the perturbation mode $k$
lasts after the beginning of contracting phase in $j$ cycle.

The amplitude of perturbation responsible for the large scale
structure of observable universe is ${\cal P}_{\zeta}^{1/2}\sim
10^{-5}$. Thus if ${\cal P}_{\zeta (i)}^{1/2}\sim 10^{-5}$ is
required in $i$ cycle, we can see when ${\cal N}_{j}\simeq {5\over
3}\ln{10}\sim 3$, ${\cal P}_{\zeta (ji )}^{1/2}\sim 1$ in $j$
cycle, where it should be noticed that when ${\cal P}_{\zeta
(ji)}^{1/2}$ approaches 1, the enhancement of nonlinear effect
will make the required ${\cal N}_{j}$ less. Thus this actually
means that nearly at the beginning time of $j$ cycle, the
curvature perturbation on super horizon scale will have the
amplitude be in order one. This will lead to the density
perturbation \be {\delta \rho\over \rho} \sim {\cal P}_{\zeta (ji
)}^{1/2}\sim 1 \ee on corresponding super horizon scale. In this
case, it is obviously impossible that the different regions of
global universe will evolve synchronously, even if they are
synchronous in $i$ cycle. This means that the global universe at
the beginning time of $j$ cycle will be inevitably separated into
many different parts, each of which actually corresponds to a new
universe and will evolve independently of one another, up to
succedent cycle \footnote{It has been argued that if a region with
super horizon scale has the density perturbation on corresponding
scale larger than 1, such a region will correspond to a separated
close universe \cite{CH}. In principle, the initial conditions of
each local universe can be obtained by rescaling the background of
parent universe. }. Therefore, it is evident that the global
universe will proliferate after each cycle. In this sense, a
cyclic multiverse actually comes into being.


However, inside each new universe, generally the amplitude of
primordial perturbation, see `b' mode in Fig.1, is contributed by
not only that of perturbations induced by the quantum fluctuation
of background field in $j$ cycle, but also that of perturbations
that enter into the horizon during the expansion of $i$ cycle and
then leave it during the contraction of $j$ cycle \footnote{We
thank R. Brandenberger for talking this point to us. }. In
general, the latter amplitude can be larger, and dependent of the
expansion behavior and the matter contents in $i$ cycle, its
spectrum is also not scale invariant. This can be seen as follows.

When $k_*\eta\simeq 1$ during $i$ cycle, $k_*$ mode just enters
into the horizon. Hereafter, its evolution obeys Eq.(\ref{uk}),
but the term ${z^{\prime\prime}\over z}$ is negligible, since
$k\gg ah$. In this case, $u_k$ is approximately constant, up to
the time when the corresponding mode leaves the horizon
\footnote{We neglected the effect of the transfer function on the
perturbation spectrum after the corresponding perturbation mode
enters into the horizon, which is dependent of the matter content
in corresponding cycle. However, the behaviors of spectrum
discussed here are not altered qualitatively}, which occurs during
the contraction in $j$ cycle. Thus we have
$|u_{i*}|=|u_{j*^{\prime}}|$, where $*^\prime$ is the time when
the corresponding mode leaves the horizon in $j$ cycle, which
implies ${\cal P}^{1/2}_{j*^\prime}=({a_{i*}\over
a_{j*^\prime}}){\cal P}^{1/2}_{i*}$. However, after this mode
leaves the horizon in $j$ cycle, its amplitude will increase like
Eq.(\ref{p3}). Thus \ba {\cal P}^{1/2}_{je} & =
&\left({h_{je}\over h_{j*^\prime}}\right){\cal
P}^{1/2}_{j*^\prime} \nonumber\\ & = & \left({k_{je}\over
k_{j*^\prime}}\right)^3\left({a_{i*}\over
a_{j*^\prime}}\right){\cal P}^{1/2}_{i*} \label{p6}\ea is
obtained, where $h\sim (ah)^3\sim k^3$ for $w\simeq 0$ has been
applied, and the subscript `$je$' denotes the end time of the
contracting phase in $j$ cycle. We can see that if ${\cal
P}^{1/2}_{i*}\sim 10^{-5}$, ${\cal P}^{1/2}_{je}$ will be larger
than that induced by the quantum fluctuation of background field
in $j$ cycle since $k_{j*^\prime}\ll k_{je}$ and $a_{i*}\simeq
a_{j*^\prime}$, and also the spectrum is quite red since ${\cal
P}^{1/2}_{je}\sim k_{j*^\prime}^{-3}$. This result implies that
there can hardly an observable universe after the bounce of $j$
cycle. In addition, it can be noticed that the amplitudes of these
modes may be also amplified up to order one, i.e. after the
contraction lasts some times, ${\cal P}^{1/2}_{j}\sim 1$. In this
case, the universe will be possibly split into smaller and smaller
fragments, which will renders the cycle ultimately end
\cite{Piao0901}, see also \cite{B0905}. Thus in this sense the
feasibility of such a cyclic multiverse scenario is questionable.

The perturbation modes that enter into the horizon during the
expansion of $i$ cycle and then leave it during the contraction of
$j$ cycle are baneful. Their amplitudes
can hardly be suppressed by certain mechanism. Thus a reasonable
solution is to push these baneful modes to larger scale, i.e.
outside of observable universe in corresponding cycle. This can be
implemented by introducing a period of accelerated expansion in
each cycle. This period can be set before the turnaround of each
cycle. In this case, there will be a period of dark energy
domination in corresponding cycle, hereafter the universe
collapses and the next cycle begins. The perturbation modes that
enter into the horizon during the expansion of $i$ cycle will
possibly leave it during the dark energy domination of $i$ cycle,
but not during the contraction of $j$ cycle. Thus the amplitude of
these modes will not increase, up to the turnaround. In this case,
${\cal P}^{1/2}_{je} = ({h_{je}\over h_{0}})({a_{i*}\over
a_0})({a_0\over a_{j*^\prime}}) {\cal P}^{1/2}_{j*^\prime}$, where
$t_0$ in Fig.2 is the time when the period of dark energy
domination begins, thus Eq.(\ref{p6}) is changed as \be {\cal
P}^{1/2}_{je} = \left({k_{je}\over {
k}_{j0}}\right)^3\left({k_{0}\over k_{i*}}\right)^3{\cal
P}^{1/2}_{i*}, \label{p8}\ee where $a\sim (ah)^{n\over n-1}\sim
k^{n\over n-1}$ has been applied, which gives that for $n={2\over
3}$, $a\sim {1\over k^2}$ and for $n\gg 1$, $a\sim k$. $k_{je}/{
k}_{j0}$ denotes the efolding number of primordial perturbation
contributed by the contraction in $j$ cycle, i.e. ${\cal
N}_{j}=\ln{({k_{je}\over {k}_{j0}})}$. The larger it is, the
larger the amplitude of perturbation is amplified during the
contraction is. This point is essentially the same as that
obtained in Eq.(\ref{p3}). While ${k_{0}/ k_{i*}}$ is the ratio of
the wavenumber of the mode that corresponds to the beginning time
of dark energy domination to that of the given mode, which
actually corresponds to the efolding number that the dark energy
phase lasts before the given mode leaves the horizon during the
dark energy domination of $i$ cycle, and is an exponentially
suppressed factor to the amplitude ${\cal P}^{1/2}_{je}$.
This result means that the period of dark energy domination
not only helps to the continuance of cycling but also the
emergence of observable universe.

\begin{figure}[t]
\begin{center}
\includegraphics[width=7.5cm]{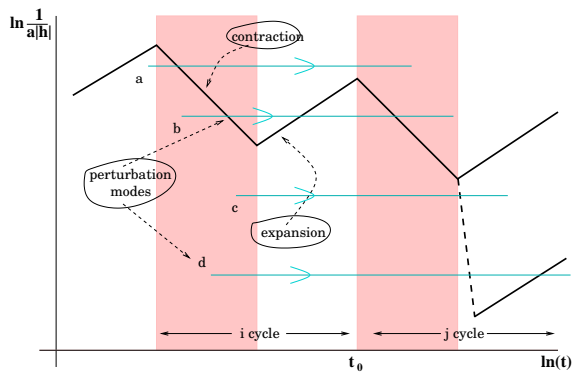}
\caption{In a cyclic universe, the sketch of $\ln{({1\over
a|h|})}$ vs. $\ln(t)$ for the case that there is a period of
inflation before bounce in $j$ cycle. The blue lines denote the
evolutions of perturbation modes, and the red regions denote the
contracting phases in each cycle. The dashed line denotes the
inflation, and the following solid line denotes the evolution
after inflation. }
\end{center}
\end{figure}

This design can be explained detailed in Fig.2. `a' mode denotes
the modes that leave the horizon in $i$ cycle but can not enter
into the horizon during the expansion of $i$ cycle. These modes
will destine to stay on super horizon scale all along up to $j$
cycle. Their amplitudes will be inevitably amplified to order one
around the beginning time of $j$ cycle. This leads to that the
universe is separated into lots of independent new universes, each
of which will evolve independently up to the succedent cycle. `b'
mode denotes the modes that enters into the horizon during the
expansion of $i$ cycle and then leaves it during the contraction
of $j$ cycle. These modes generally have amplitudes larger than
that induced only by the fluctuation of background field in $j$
cycle, which thus is baneful for the existence of observable
universe in $j$ cycle. The dashed line denotes the period of dark
energy domination, which can push those baneful modes, such as `b'
mode, outside of observable universe. In this case, `d' mode will
provide the primordial perturbation of observable universe, which
is induced only by the quantum fluctuation of background field in
$j$ cycle, and its amplitude is given by (\ref{p1}) and thus can
be suitable for observations \footnote{The inflation might occur
after the bounce in some cycles, see Fig.3, since the energy scale
of bounce is generally high, it can be expected that after the
bounce the field might land on a higher plain of effective
potential. In this case, `b' mode will be pushed outside of
observable universe by the inflation in $j$ cycle, and the
primordial perturbations suitable for observable universe, such as
`c' and `d' modes, can emerge during inflation, which are induced
by the quantum fluctuation of inflaton field. The inflation after
bounce has been originally studied in Refs. \cite{PFZ, PTZ}, in
which the imprint of bounce on cosmic microwave background has
been pointed out, see also for latest studies \cite{M},\cite{CZ}.
Here for cyclic multiverse, the occurrence of inflation has
similar effect as that of a period of dark energy domination.
However, both can be different in the theoretical model building
and observational signals, which might be interesting for
studying.}.

The causal patch diagram of cyclic multiverse can be plotted in
Fig.4, which is obtained by gluing the corresponding parts of
diagrams of contracting phase, expanding phase and dS phase. This
causal diagram is slightly similar to that of inflationary
multiverse, e.g.\cite{Wini}. The reason is that before the
turnaround in each cycle the universe is in a dS phase. In some
sense, such a period leads that the universe has an average
positive energy density for each cycle. In inflationary
multiverse, the new universe is generated either in nucleated
bubble, e.g. \cite{G},\cite{Steinhardt83}, or in the region that
the inflaton field is in stochastic walking
\cite{vilenkin},\cite{linde}, both are induced by quantum effects
\footnote{However, in slow roll inflation, the multiverse can be
also induced by the classical rolling of inflaton along a web of
branches of its effective potential \cite{LLP}.
}. Here, however,
the mechanism resulting in multiverse is the amplification of the
amplitude of curvature perturbation cycle by cycle, which thus
occurs in classical sense, though it initially originates from the
quantum fluctuation of field.

\begin{figure}[t]
\begin{center}
\includegraphics[width=6.5cm]{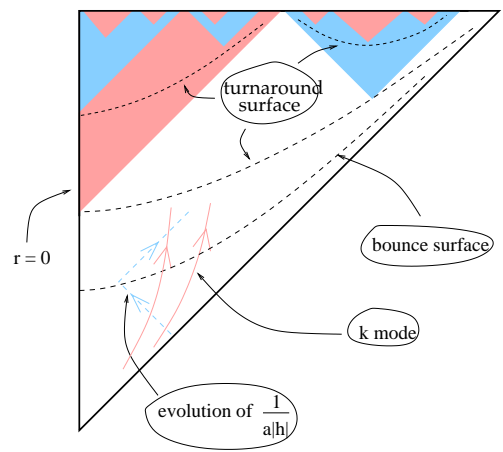}
\caption{The Penrose diagram of a cyclic multiverse. This is
obtained by gluing the corresponding parts of diagrams of
contracting universe, expanding universe and dS universe, where
the dashed lines denoting the turnaround surface and the bounce
surface have been signed respectively. Before each turnaround
surface, the universe will experience the contraction, bounce,
expansion successively and then enter into a period of dark energy
domination. After this, the parent universe is separated into lots
of parts independent of one another, each of which corresponds to
a new universe and repeats above evolution. In principle, since
the experience of each universe after the proliferation is
generally different, they will not be expected to be synchronous
in cycling. This means that generally when we are in a period of
dark energy domination, it is possible that there are many other
universes which are in the period of contraction or bounce or
others.
 }
\end{center}
\end{figure}



We can have a concrete implement to this multiverse scenario, like
in \cite{Piao04}. In this implement, the potential of scalar field
has a nearly flat region, the value of whose energy density equals
to that of cosmological constant observed, and a minimum with
negative energy density. When the field is in the nearly flat
region of potential, we can have a period of dark energy, and then
it rolls down towards its minimum with negative potential energy
density, which leads to the collapse of universe, the contraction
with $w\simeq 0$ can be obtained by the oscillation of field
around its minimum. In this case, it is obvious that the evolution
that the period of dark energy domination precedes the contraction
with $w\simeq 0$ can be obtained. In general, the period of dark
energy will help to dilute the matter and radiation in $i$ cycle,
which assures that in $j$ cycle the energy density of matter and
radiation from $i$ cycle can not exceed that of field till the
enough efolding number is obtained. In addition, the anisotropy
and some baneful leftovers, which are menaces for cycling, e.g.
discussions in \cite{BD}, can be also diluted during this period
of dark energy. Thus it is required that this period should be
enough long. i.e. the corresponding potential should be enough
flat.

In different cycle of cyclic universe, the universe can be in
different minima of a given potential in field space, or landscape
\footnote{In the low energy limit, the string landscape can be
visualised as an effective potential in a given field space with
multiple dimensions.} \cite{Piao04}, in which these minima have
negative potential energy density. Here, the generality is
actually straight. Thus in principle it can be argued that in a
cyclic multiverse scenario, generally each of multiverse will have
different minima and thus evolutions. However, in this case, the
requirement for an enough period of accelerated expansion in $i$
cycle actually corresponds to a fine tunning for the initial
condition of $j$ cycle.
Thus though in a cyclic universe driven by a given landscape, the
number of universes having an enough long period of dark energy
domination might be quite small, such universes can certainly
exist, which is significant not only for assuring the continuance
of cycle, but also for the emergence of observable universe in
which we might live. In general, in a string landscape both
positive and negative minima exist. However, unless all minima in
the landscape are negative, the cycle of universe will not
continue eternally, since if the universe enters into a positive
minimum, it will stop cycling, and expand for ever, as illustrated
earlier in \cite{Biswas1}.


In cyclic universe \cite{STS}, the effect of the increase of
metric perturbation on global universe has been discussed in
\cite{Erick}. However, in this model it requires that on super
horizon scale the increasing mode of metric perturbation is
inherited by the constant mode of curvature perturbation in
leading order. Whether the corresponding inheriting can occur
remains controversial
\cite{Lyth},\cite{DV},\cite{TBF},\cite{GKST},\cite{PZ11},\cite{BV},\cite{ABB},\cite{CQB}.
However, if this dose not occur, the same mode is decayed during
the expansion of each cycle, thus even if it is increased during
the contraction, the net result is still decayed in each cycle,
since in each cycle the amount of the expansion of the universe is
generally larger than that of the contraction. Here, however, the
increasing mode of metric perturbation is inherited by the
constant mode of curvature perturbation in $k^2$ order, which
certainly occurs for the bounce connecting the contracting and
expanding phases, e.g. \cite{AW, Cai0810} for theoretical and
numerical studies. The metric perturbation after the bounce will
be dominated by the same constant mode as that of curvature
perturbation. Thus the net amplitude of metric perturbation is
increased in each cycle. We has argued this effect will
potentially lead to a cyclic multiverse \cite{Piao0901}.
In this paper, the problems baffling this multiverse scenario is
discussed, and the possible solution is designed.

In a class of oscillating universe, in which the bounce is
implemented by quintom matter
\cite{Xiong},\cite{Cai0704},\cite{FBK}, the cyclic multiverse
scenario might be naturally applied, since a period of dark energy
can be congenitally included in this model. This study can be in
order.
In \cite{BF}, a scenario that many small contracting universes can
be spawned at the end of each cycle has been proposed, which is
implemented by appealing to the brane world cosmology and the
phantom dark energy. In the design given here, these additional
appeals are needless, the mechanism resulting in the multiverse is
the natural increase of perturbation on super horizon scale. The
role of a period of dark energy for cycling was also discussed in
\cite{GD}, in which the appearance of new universe is obtained by
the wormholes leaded by the accretion of phantom dark energy
\cite{GD1}.

In cyclic inflation \cite{Biswas},\cite{Biswas1}, the scale factor
in consecutive cycles increases by a constant factor. This, after
some cycles, can give a net exponential growth of the scale
factor, and thus can imitate inflation. Whether there can be a
scale invariant spectrum in this model remains an open issue.
However, combining it with the result obtained here, we might
argue that a multiverse will come into being after cyclic
inflation, some of which might have a subsequent period of slow
roll inflation and thus correspond to our observable universe.


In conclusion, in a cyclic universe, since the metric perturbation
on super horizon scale is amplified cycle by cycle, after each
cycle the universe will be inevitably separated into many parts
independent of one another, each of which corresponds to a new
universe and evolves up to succedent cycle, and then is separated
again. This mechanism brings us a scenario of cyclic multiverse,
in which the number of universes increases cycle by cycle.
However, in general, the amplitude of perturbation modes in
previous cycle can be preserved and amplified in current cycle,
which is generally larger than that induced by the quantum
fluctuation of background field in current cycle. This baffles the
possibility that this scenario is regarded as the origin of
observable universe. We, in this paper, have provided a viable
design of a cyclic multiverse, in which the observable universe
can emerge naturally.
The significance of a long period of dark energy before the
turnaround of each cycle for this implementing is shown.
In this design, the causal diagram of cyclic multiverse likes
that of inflationary multiverse. Thus the measure for the
multiverse might be discussed similarly.
Dark energy is an important issue of current cosmology, which is
being intensively explored. In some sense, this work might provide
an alternative motivation for the existence of dark energy.




\textbf{Acknowledgments} We thank R. Brandenberger, Y.F. Cai, J.
Zhang for helpful discussions. This work is supported in part by
NSFC under Grant No:10775180, in part by the Scientific Research
Fund of GUCAS(NO:055101BM03),
in part by National Basic Research Program of China,
No:2010CB832805



\begin{thebibliography}{99}


\bibitem{Tolman} R.C. Tolman, Relativity, Thermodynamics and Cosmology, (Oxford U. Press,
Clarendon Press, 1934).

\bibitem{STS} P.J. Steinhardt, N. Turok, Science \textbf{296},
(2002) 1436; Phys. Rev. \textbf{D65} 126003 (2002).







\bibitem{BD} J. Barrow, M. P. Dabrowski, Mon. Not. R. Astr. Soc. \textbf{275}, 850
(1995).

\bibitem{KSS} N. Kanekar, V. Sahni, Y. Shtanov, Phys. Rev. \textbf{D63}, 083520
(2001).

\bibitem{Piao04} Y.S. Piao, Phys. Rev. \textbf{D70}, 101302 (2004);
Y.S. Piao, Y.Z. Zhang, Nucl. Phys. \textbf{B725}, 265 (2005).



\bibitem{Lidsey04} J.E. Lidsey, D.J. Mulryne, N.J. Nunes, R. Tavakol,
Phys. Rev. \textbf{D70}, 063521 (2004).





\bibitem{CB} T. Clifton, J.D. Barrow, Phys. Rev. \textbf{D75}, 043515 (2007).

\bibitem{Xiong} H.H. Xiong, Y.F. Cai, T. Qiu, Y.S. Piao, X.M.
Zhang, Phys. Lett. \textbf{B666}, 212 (2008).


\bibitem{Xin} X. Zhang, Eur. Phys. J. \textbf{C60}, 661 (2009).

\bibitem{LS} J.L. Lehners, P.J. Steinhardt, Phys. Rev. \textbf{D79}, 063503 (2009).


\bibitem{Biswas} T. Biswas, S. Alexander, Phys. Rev. \textbf{D80}, 043511 (2009).

\bibitem{Biswas1} T. Biswas, A. Mazumdar, Phys. Rev. \textbf{D80}, 023519
(2009).

\bibitem{NB} M. Novello, S.E.P. Bergliaffa, Phys. Rept. \textbf{463}, 127 (2008).



\bibitem{Piao0901} Y.S. Piao, Phys. Lett. \textbf{B677}, 1 (2009).

\bibitem{Feng04} B. Feng, M.Z. Li, Y.S. Piao, X.M. Zhang, Phys. Lett. B634,
101 (2006).

\bibitem{IBF} C. Ilie, T. Biswas, K. Freese. arXiv:0908.0991.

\bibitem{PZ} Y.S. Piao, E Zhou, Phys. Rev. \textbf{D68}, 083515
(2003).

\bibitem{PZ1} Y.S. Piao, Y.Z. Zhang, Phys. Rev. \textbf{D70},
063513 (2004).

\bibitem{BKM} T. Biswas, T. Koivisto, A. Mazumdar,
arXiv:1005.0590.

\bibitem{Cai0704} Y.F. Cai, T. Qiu, Y.S. Piao, M.Z. Li, X.M.
Zhang, JHEP \textbf{0710}, 071 (2007).

\bibitem{AJK} I.Y. Arefeva, L.V. Joukovskaya, A.S. Koshelev, JHEP
\textbf{0309}, 012 (2003).

\bibitem{AKV} I.Y. Arefeva, A.S. Koshelev, S.Y. Vernov, Phys. Rev.
\textbf{D72}, 064017 (2005).



\bibitem{Muk} V.F. Mukhanov, JETP lett. 41, 493 (1985); Sov. Phys. JETP. 68,
1297 (1988).

\bibitem{KS} H. Kodama, M. Sasaki, Prog. Theor. Phys.
Suppl. 78 1 (1984).

\bibitem{Wands99} D. Wands, Phys. Rev. \textbf{D60}, 023507
(1999).

\bibitem{FB} R. Brandenberger, F. Finelli, JHEP \textbf{0111} (2001) 056;
F. Finelli, R. Brandenberger, Phys. Rev. \textbf{D65}, 103522
(2002).


\bibitem{PP} P. Peter, N. Pinto-Neto, Phys. Rev. \textbf{D78}, 063506 (2008).

\bibitem{Wands09} D. Wands, arXiv:0809.4556.

\bibitem{Cai0810}  Y.F.Cai, T. Qiu, R. Brandenberger, X.M. Zhang,
Phys. Rev. \textbf{D80}, 023511 (2009).

\bibitem{S} A.A. Starobinsky, JETP Lett. \textbf{30}, 682 (1979).



\bibitem{CH} B. Carr, S.W. Hawking, Mon. Not. R. Astro. Soc. \textbf{168},
399 (1974).

\bibitem{B0905} R. Brandenberger, Phys. Rev. \textbf{D80}, 023535 (2009).


\bibitem{PFZ} Y.S. Piao, B. Feng, X.M. Zhang, Phys. Rev. \textbf{D69},
103520 (2004).

\bibitem{PTZ} Y.S. Piao, S. Tsujikawa, X.M. Zhang, Class. Quant. Grav. \textbf{21}, 4455 (2004).

\bibitem{M} J. Mielczarek, JCAP \textbf{0811}, 011 (2008);
arXiv:0908.4329.

\bibitem{CZ} Y.F. Cai, X.M. Zhang, JCAP \textbf{0906}, 003 (2009).

\bibitem{Wini} S. Winitzki, Lect. Notes Phys. \textbf{738}, 157
(2008).

\bibitem{G} J.R. Gott, Nature \textbf{295}, 304 (1982).

\bibitem{Steinhardt83} P.J. Steinhardt, in ``The Very Early Universe" ed. by
G.W. Gibbons, S.W. Hawking and S. T. C. Siklos (Cambridge
University Press, Cambridge, 1983).

\bibitem{vilenkin} A. Vilenkin, Phys. Rev. \textbf{D27}, 2848 (1983).

\bibitem{linde} A. Linde, Phys. Lett. \textbf{B175}, 395 (1986).

\bibitem{LLP} S. Li, Y. Liu, Y.S. Piao, Phys. Rev. \textbf{D80},
123535 (2009).




\bibitem{Erick} J.K. Erickson, S. Gratton, P.J. Steinhardt, N.
Turok, Phys. Rev. \textbf{D75}, 123507 (2007).


\bibitem{Lyth} D.H. Lyth, Phys. Lett. B524, 1 (2002); Phys. Lett. B 526, 173
(2002).

\bibitem{DV} R. Durrer, F. Vernizzi, Phys. Rev. D66, 083503 (2002).

\bibitem{TBF} S. Tsujikawa, R. Brandenberger, F. Finelli, Phys. Rev. D66,
083513 (2002).

\bibitem{GKST} S. Gratton, J. Khoury, P.J. Steinhardt, N. Turok, Phys. Rev.
D69, 103505 (2004).

\bibitem{PZ11} Y.S. Piao, Y.Z. Zhang, Phys. Rev. D70 043516 (2004).

\bibitem{BV} V. Bozza, G. Veneziano, JCAP 009, 005 (2005); V. Bozza, JCAP
0602, 009 (2006).

\bibitem{ABB} S. Alexander, T. Biswas, R.H. Brandenberger,
arXiv:0707.4679.

\bibitem{CQB} Y.F. Cai, T. Qiu, R. Brandenberger, Y.S. Piao, X.M. Zhang,
JCAP 0803, 013 (2008).

\bibitem{AW} L.E. Allen, D. Wands, Phys. Rev. D70, 063515 (2004).



\bibitem{FBK} K. Freese, M.G. Brown, W.H. Kinney, arXiv:0802.2583.

\bibitem{BF} L. Baum and P.H. Frampton, Phys. Rev. Lett. \textbf{98},
071301 (2007).

\bibitem{GD} P.F. Gonzalez-Diaz, arXiv:0908.3244.

\bibitem{GD1} P.F. Gonzalez-Diaz, Phys. Rev. Lett. \textbf{93},
071301 (2004); P.F. Gonzalez-Diaz, J.A. Jimenez-Madrid, Phys.
Lett. \textbf{B596}, 16 (2004).

















\end{thebibliography}
\end{document}